# Deciphering the underlying mechanisms of the pharyngeal motions in *Caenorhabditis elegans*

**Dana Sherman and David Harel**

(Department of Computer Science and Applied Mathematics, Weizmann Institute of Science, Israel)

The pharynx of the nematode *Caenorhabditis elegans* is a neuromuscular pump that exhibits two typical motions: pumping and peristalsis. While the dynamics of these motions are well characterized, the underlying mechanisms generating most of them are not known. In this paper, we propose comprehensive and detailed mechanisms that can explain the various observed dynamics of the different pharyngeal areas: the dynamics of the pumping muscles – corpus, anterior isthmus, and terminal bulb – and the peristalsis dynamics of the posterior isthmus muscles. While the suggested mechanisms are consistent with all available relevant data, the assumptions on which they are based and the open questions they raise could point at additional interesting research directions on the *C. elegans* pharynx. We are hoping that appropriate experiments on the nematode will eventually corroborate our results, and improve our understanding of the functioning of the *C. elegans* pharynx, and possibly of the mammalian digestive system.

## Keywords

# Introduction

The pharynx of the nematode *C. elegans* is a double-bulbed tube, composed of 20 muscle cells [and of several other types of cells; Fig 1; 1]. Anatomically, these 20 muscle cells can be divided into 8 types, due to tri-diagonal symmetry of most of them around the centre of the pharynx. Functionally, the 20 muscle cells can be divided into three groups, such that within each group all muscles contract and relax in synchrony, as if they were a single muscle cell: (a) *corpus* muscles: pharyngeal muscles 1-4 (pm1-4; pm1-2 actually do not contract in synchrony with pm3-4; still pm1-4 are classified as corpus muscles), (b) *isthmus* muscles: pm5, and (c) *terminal bulb* muscles: pm6-8.

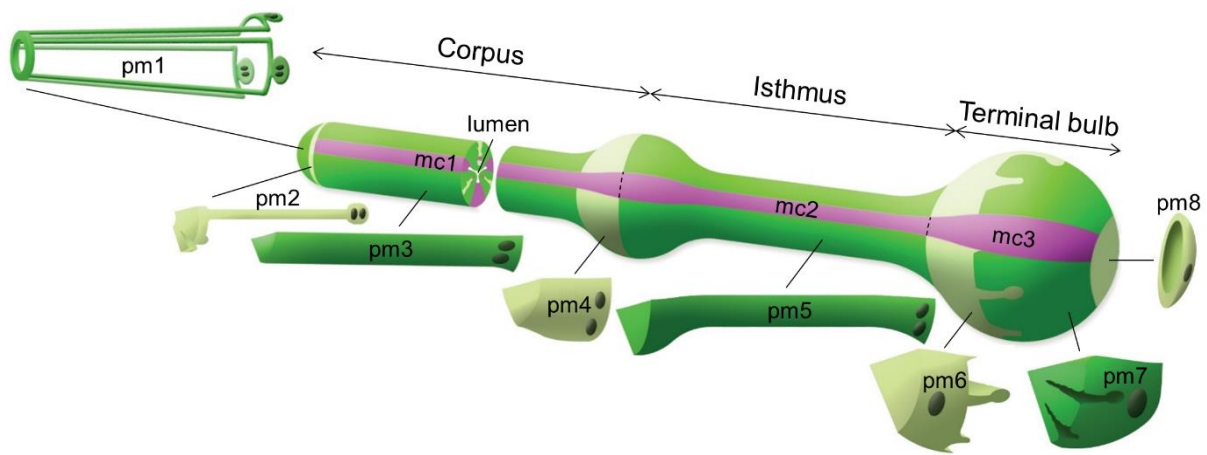

**Fig 1. The anatomy of *C. elegans* pharynx.** In all figures, anterior is to the left and dorsal is up. The pharynx is composed of several types of cells, including 20 muscle cells (greens), 9 structural cells (named marginal cells; purple), 20 neurons, 9 epithelial cells, and 4 gland cells (not shown). All types of pharyngeal muscle cells (pm), except pm1 and pm8, and all types of marginal cells (mc) have three copies arranged with three-fold symmetry around the pharyngeal lumen. Pharyngeal muscles 1-4 (pm1-4), as well as marginal cells 1 (mc1), compose the anterior part of the pharynx – the corpus, pm5 and mc2 compose its middle part – the isthmus, and pm6-8, together with mc3, compose its posterior part – the terminal bulb. Adapted from Altun and Hall [1].

All pharyngeal muscles, except pm1-2, exhibit one of two typical motions: *pumping* or *peristalsis* (Fig 2). Pumping is a repetitive contraction and relaxation cycle of most of the organ, in a way resembling vertebrate heart-beats. Pumping of the anterior pharynx, i.e.,

pm3-4 and anterior part of pm5, sucks-in bacterial food from the environment, and pumping of the posterior part, i.e., pm6-8, crushes the food and pushes it into the intestine. Peristalsis is an anterior-to-posterior wave of local contractions and relaxations, one after the other, of the non-pumping pharyngeal segment, i.e., the posterior part of pm5. It occurs once in every 3-4 pumps on average, at the end of a pump cycle [2]. The separation into two motions allows a more efficient ingestion of food, in which peristalsis transports food against pressure differentials – from low environmental pressure, at the anterior pharynx, to high pressure at its posterior end [3].

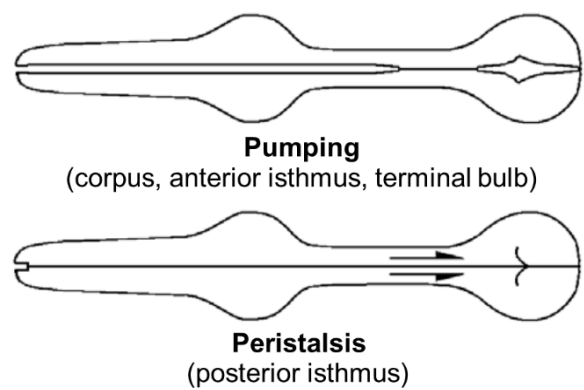

**Fig 2. The typical motions of *C. elegans* pharynx: pumping and peristalsis.** (**Top**) Pumping consists of cycles of simultaneous contractions and relaxations of entire muscle groups, which open and close the pharyngeal lumen. In *C. elegans*, pumping occurs in the corpus, anterior isthmus, and terminal bulb (indicated by the open lumen). (**Bottom**) Peristalses are posteriorly moving contraction waves, occurring in the posterior isthmus in *C. elegans* (indicated by arrows). Adapted from Chiang, Steciuk [4].

Inspecting pumping dynamics in detail reveals a more complicated picture than perfectly synchronized contractions and relaxations of the various pharyngeal areas [Fig 3; Table 1; 3, 5, 6-8]. Contraction dynamics of the pumping areas differ in three aspects: (1) contraction-onset time, (2) contraction-spreading speed, and (3) contraction-strength progression. Relaxation dynamics of the pumping areas differ in relaxation-onset time. The slightly different contraction and relaxation times of the corpus and anterior isthmus allow for more efficient transport of food along the anterior pharynx [7, 8]. Posterior isthmus motion is also well characterized [Fig 3; Table 1; 2, 3].

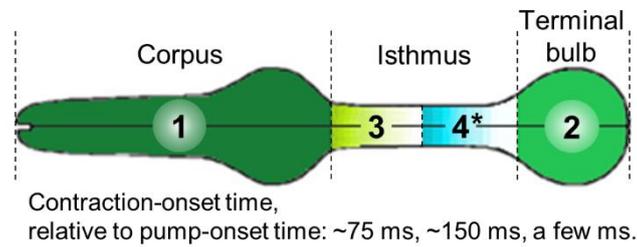

**Fig 3. Contraction dynamics of the various pharyngeal areas.** Contraction dynamics vary between different pharyngeal muscle groups, where the corpus and terminal bulb muscles contract very early during the pump, simultaneously along their entire length (solid fills), while the contraction of the isthmus muscles is delayed and spreads slowly in an anterior-to-posterior wave within several tens of milliseconds (gradient fills). Another difference is between the contraction-strength progression of the corpus and terminal bulb muscles, where the corpus contracts gradually throughout the pump – weakly at the beginning of a pump and reaching maximum contraction towards its end – while the terminal bulb reaches maximum contraction very early after pump-onset, and remains at maximum contraction throughout the pump (not shown). Contraction order and the explicit delays are indicated on and under the figure, respectively. The pumping and peristaltic muscles are colored in greens and cyan, respectively. *Posterior isthmus peristalsis occurs once per 3-4 pumps, on average.

**Table 1. Contraction and relaxation dynamics of the various pharyngeal areas.**

| Motion | Feature | Corpus | Anterior isthmus | Posterior isthmus | Terminal bulb |
|---|---|---|---|---|---|
| Contraction | Onset time[a] [ms] | 0 | ~75 | ~150 ms[c] | A few milliseconds |
| | Spreading speed | Rapid | Slow, within ~85 ms | Slow | Rapid |
| | Strength progression | Gradual | - | - | Rapid |
| Relaxation | Onset time[b] [ms] | 0 | ~20 | | [d] |
| | Spreading speed | Rapid | Rapid | Slow | Rapid |

Note that while the corpus, anterior isthmus and terminal bulb pump, posterior isthmus exhibits peristalsis.

[a] Relative to contraction-onset time of the corpus, defined as "pump-onset time".

[b] Relative to relaxation-onset time of the corpus, defined as "relaxation-onset time".

[c] Does not occur after every pump, but once per 3-4 pumps, on average.

[d] Terminal bulb's repolarization usually occurs less than 50 ms after that of the corpus [9]. However, no data is available for the time delay between repolarization and relaxation for neither the corpus nor the terminal bulb. Thus, we have no estimation of the relaxations time-difference between the two areas.

While the dynamics of the two typical pharyngeal motions are well characterized, the underlying mechanisms generating most of them are not known. (As far as we know, explanations for several aspects of the dynamics exist [e.g., 2, 5, 10], but these are partial, either not explaining all relevant data or not explaining the observed phenomena in sufficient depth.) Several questions arise regarding the way the different contraction and relaxation dynamics of the various pharyngeal areas could occur, where the most interesting ones involve isthmus dynamics: not only does this pharyngeal segment, composed of a single type of muscle cells, exhibit two different motions that occur at different times and frequencies (which is partially explained by Song and Avery [2]), but its contraction dynamics are fundamentally different from those of all other pharyngeal areas: late contraction-onset times and slowly-spreading contractions. These dynamics are puzzling, since they do not accord with the dynamics of the underlying processes that lead to pharyngeal muscle contraction: depolarization (DP) of the muscles, which induces elevation of intracellular calcium ions, which in turn induces muscle contraction. Specifically, as most data are available for the easier-to-measure DP process, the delayed and slow contraction of the isthmus breaks the rapid middle-to-edges spread of DP along the organ [Fig 4; 3].

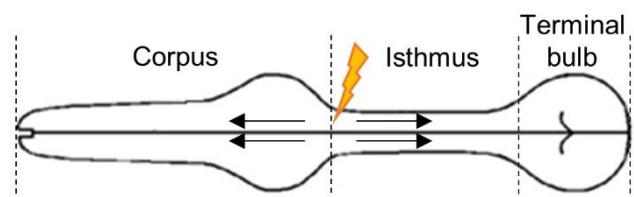

**Fig 4. Depolarization dynamics of *C. elegans* pharyngeal muscles.** Depolarization is initiated by the firings of a pair of pharyngeal neurons (MC's) that innervate the corpus-isthmus border [11-13]. Their firings induce an electrical signal that is generated locally at the innervation site (lightning), and which actively propagates along the pharynx within a few milliseconds (arrows) [3, 14].

In this paper, we propose comprehensive, detailed mechanisms that explain the various measured dynamics of all pharyngeal areas. The first section describes our thinking process for trying to decipher the most puzzling open questions, regarding isthmus contraction dynamics. Since more quantitative data is available for anterior isthmus (AI) pumping, we start by focusing on this region. Out of several hypotheses with which we were able to come up, only one matched all known data. This hypothesis set the ground for the proposed mechanism of AI dynamics. In the second section, we propose possible mechanisms for the observed dynamics of the other pharyngeal parts – posterior isthmus (PI), corpus, and terminal bulb (TB) – based on our hypothesized mechanism for the AI.

# Deciphering the contraction dynamics of the anterior isthmus muscles

As mentioned, we started by trying to tackle the fundamentally different AI contraction dynamics. Since three processes underlie pharyngeal muscle contraction, we assumed that the gradual contraction of the AI results from slow dynamics of at least one of these processes. We started by assuming slow dynamics of the first process – DP along the AI muscle (hypothesis #1). Contradictory findings led us to rule out this option and to examine the next possible process – calcium dynamics (hypothesis #2). Also here, experimental findings did not support slow calcium dynamics, forcing us to conceive of another mechanism that would accord with all available relevant data as described below (hypothesis #3).

**Hypothesis #1 for the *gradual* contraction of the anterior isthmus: *passive* conductance of the electrical signal along the anterior isthmus muscles**

As already mentioned, there is a large time-difference between the rapidly-spreading DP along the entire pharynx and the slowly-spreading contraction along the isthmus. The former results from two reasons: (1) active propagation of the electrical signal along the pharyngeal muscle cells [15, 16], and (2) strong electrical coupling between adjacent muscle cells, via gap junctions [1].

Avery and Thomas [3] suggested that "the lack of synchrony in isthmus contraction can be explained by proposing that isthmus muscle, unlike TB or corpus muscle, is incapable of regenerative action potentials. In this case, local excitation of the muscle would produce a local DP and local contraction, both of which would tend to spread slowly from the site of excitation and decrease with distance. Thus, the delayed contraction of the AI during a pump would be explained by excitation at the anterior end through electrical coupling to corpus muscle cells."

Assuming passive conduction along the isthmus muscles following the hypothesis of Avery and Thomas [3] means that the TB could depolarize within a few milliseconds after the corpus only if the corpus-TB electrical coupling bypassed the isthmus muscles. This could occur in two ways (Fig 5A):

(1) Via pharyngeal neurons: The pharynx is innervated by 20 neurons, divided into 14 types due to right-left symmetry of some of them, with many stretching between the corpus-isthmus and isthmus-TB borders: the M2's pair, M4, I4, I5, I6, and the NSM's pair [11]. While anatomically such a bypass is possible, several pieces of evidence do not accord with this idea: (a) no gap junctions exist between the pharyngeal muscles and neurons, where only such connections could explain the rapid spreading [1]; (b) even when killing all pharyngeal neurons, corpus and TB still contract in synchrony, suggesting that the neurons are not required for the tight coupling [17].

(2) Via marginal cells: As shown in Fig 1, three types of marginal cells stretch along the pharynx: mc1 along the corpus, mc2 along the isthmus, and mc3 along the TB. Thus, the mc2, which directly connect the corpus-isthmus and isthmus-TB borders, are anatomically good candidates for bypassing the isthmus muscles. In addition, the mc2 are connected to the pharyngeal muscles via gap junctions: within each pharyngeal area, the marginal cells are linked to neighbouring muscle cells by gap junctions [3]; between pharyngeal areas, the marginal cells form gap junctions to neighbouring marginal cells [1]. Thus, the mc2 could couple the corpus-TB muscles via the following gap-junctions route: corpus muscles → mc1 → mc2 → mc3 → TB muscles. Since signal delay in *C. elegans* gap junction is ~0.2 ms, the total delay sums up to less than 1 ms, as desired [18]. However, the idea of an mc2-bypass does not accord with the following findings: (a) dye injected to the corpus muscles spreads through the isthmus muscles to the TB, not entering the mc2 [19]; (b) the gap junctions in different pharyngeal cells are encoded by different genes. In mutants in which the gap-junction genes of the mc2, which are also expresses in the corpus muscles, are not expressed, pm4 and TB still contract in synchrony, just as in the intact worm (whereas pm3 contractions occur separately and independently), suggesting that the mc2 are not required for the tight coupling [6].

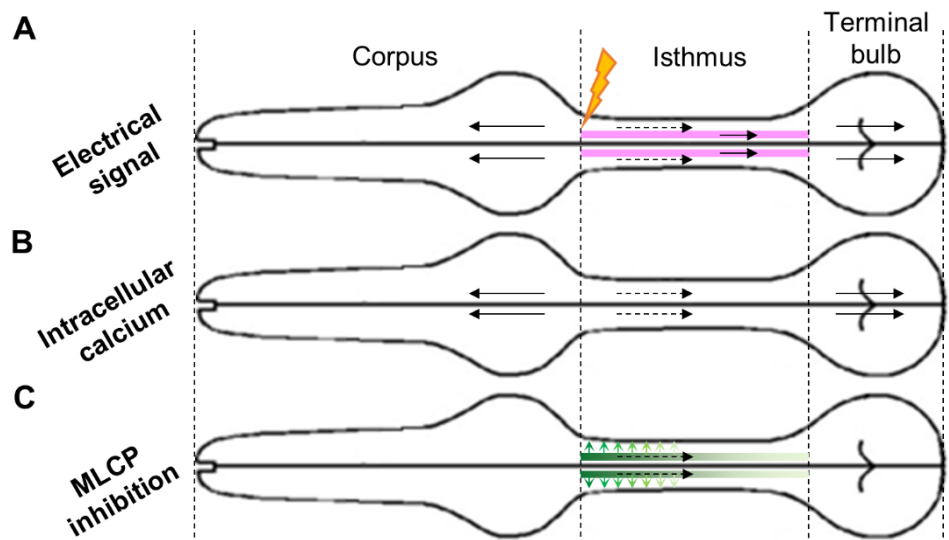

**Fig 5. Hypotheses for the *gradual* contraction of the anterior isthmus.** (**A**) Hypothesis #1, ruled out: passive conductance of the electrical signal along the AI muscles**.** The locally-initiated DP of the pharyngeal muscles at the corpus-isthmus border by the MC's firings (lightning), propagates rapidly (slowly) along the corpus and TB (isthmus) muscles, due to active (passive) conductance (solid (dashed) arrows); it is also conducted rapidly (solid

arrows) along a bypass of non-muscular cells stretching throughout the isthmus (pink). (**B**) Hypothesis #2, ruled out: slow spread of calcium ions along the AI muscles. In contrast to the corpus and TB muscles, in which the rapidly-spreading DP induces rapid, nearly-simultaneous elevation of intracellular calcium ions ($[Ca^{+2}]_{in}\uparrow$) throughout their lengths (solid arrows), DP of the isthmus muscles does not induce $[Ca^{+2}]_{in}\uparrow$. Alternatively, local $[Ca^{+2}]_{in}\uparrow$ at the corpus-isthmus border, following its innervation by pharyngeal neurons, e.g., MI or MC's, results in a slow spreading towards the TB by diffusion (dashed arrows). (**C**) Hypothesis #3, possible: gradual inhibition of a muscle-contraction regulating enzyme, myosin-light-chain-phosphatase (MLCP), along the AI muscles. Contraction of the non-striated pharyngeal muscles depends on the balance between the activity-level of two regulatory enzymes: myosin-light-chain-kinase (MLCK) and MLCP. In the relaxed isthmus muscles, MLCP activity is high, and thus its inhibition, in addition to MLCK-activation by DP, is required for muscle contraction. MLCP-inhibition occurs gradually along these muscles, due to a slow anterior-to-posterior conductance of the electrical signal (dashed arrows) along the muscles' effectors (green). These apply many innervations throughout the isthmus, such that each induces local inhibition of MLCP at the adjacent muscles' segments (green arrows).

The above do not support the idea of bypassing the electrical conductance of the isthmus muscles via other pharyngeal cells, and thus imply active conductance along the isthmus muscles. Several findings support the latter, with the strongest one showing that mutants in which the gap-junction genes of the pm4 and isthmus muscles are not expressed have uncoupled contractions of the corpus and TB muscles [19, 20]. This implies that the isthmus muscles are required for the rapid relay of the electrical signal, resulting in coordinated corpus-TB muscle contraction.

Thus, in the following hypotheses, we assume a rapid DP of the isthmus muscles, which occurs before TB's DP.

**Hypothesis #2 for the *gradual* contraction of the anterior isthmus: slow spread of calcium ions along the anterior isthmus muscles**

The next process that could explain the lagging contraction dynamics of the isthmus muscles is calcium ($Ca^{+2}$) dynamics. The electrical signal generated along the pharyngeal muscles results from the sequential opening and closing of several types of ion channels, including $Ca^{+2}$ channels [16]. In *C. elegans*, the latter include two types, encoded by two genes, *cca-1* and *egl-19*, where the CCA-1 channels initiate the DP process, and EGL-19 channels maintain it. However, different ion channels could contribute differently to the generation of the electrical signal at different pharyngeal muscle cells, generating varying-amplitude currents; for example, due to unevenly-distributed expressions [e.g., 21]. Specifically, the expression level of the EGL-19 channels, which are the main $Ca^{+2}$ channels responsible for DP and $Ca^{+2}$ influx in the pharyngeal muscles [16, 22, 23], could be low at the isthmus muscles, not allowing a large increase in intracellular calcium ions ($[Ca^{+2}]_{in}\uparrow$) during most of the DP. In such a case, local innervation of the corpus-isthmus border (for example, by the pharyngeal neuron MI or by the MC's pair), could induce local $[Ca^{+2}]_{in}\uparrow$, which could spread slowly along the isthmus muscles by diffusion, inducing gradual muscle contraction [Fig 5B; 11].

We thus calculated the time it would take $Ca^{+2}$ to diffuse along the AI muscle, as follows: The time τ it would take molecules to traverse a linear distance R by diffusion in a cell, given their diffusion coefficient in cytoplasm, D, is $τ=R^2/6D$ [24]. $Ca^{+2}$ diffusion coefficient in cytoplasm is D = 0.53 $μm^2$/ms, and the length of the AI is ⅓-½ of the entire isthmus length, i.e., 11.93-17.9 μm, resulting in τ = 44.78-100.76 ms [7, 25, 26]. These values fit very well the measured ~85 ms its takes the contraction wave to spread along the AI muscles.

However, the findings in Shimozono, Fukano [27], who directly measured $Ca^{+2}$ dynamics from pharyngeal muscles, show that $Ca^{+2}$ dynamics at the AI are similar to those at the TB; i.e., they tightly follow the DP. In addition to the very early $[Ca^{+2}]_{in}\uparrow$ at the beginning of a pump, the measurements from a relatively posterior position along the AI indicate a nearly simultaneous rise throughout this entire segment, in contrast to our hypothesized slow spread of $Ca^{+2}$ along the AI, which caused us to reject this hypothesis.

The early [Ca$^{+2}$]$_{in}$↑ does not accord with the late isthmus contraction's observation. We therefore temporarily stopped tackling the slowly-spreading contraction question and turned to focus on the large contraction-onset delay.

**Hypothesis #1 for the *delayed* contraction of the anterior isthmus: delayed inhibition of a muscle-contraction regulatory enzyme – myosin-light-chain-phosphatase (MLCP)**

Two types of muscles exist in most organisms, whose contraction depends on a different set of conditions (Fig 6):

(1) *Striated muscles*, in which myosin-actin cross-bridges, which induce contraction, are formed only when the myosin binding sites on the actin filaments are exposed. The binding-sites exposure is a Ca$^{+2}$-dependent process, and thus, in striated muscles, [Ca$^{+2}$]$_{in}$↑ is required and sufficient for inducing muscle contraction (Fig 6A).

(2) *Non-striated muscles*, in which myosin-actin cross-bridges are formed only when the myosin heads are phosphorylated. While the myosin-heads phosphorylation is a Ca$^{+2}$-dependent process, [Ca$^{+2}$]$_{in}$↑ is required but is not sufficient for non-striated muscle contraction, due to a competing process that occurs in parallel to the phosphorylation – myosin-heads de-phosphorylation (Fig 6B).

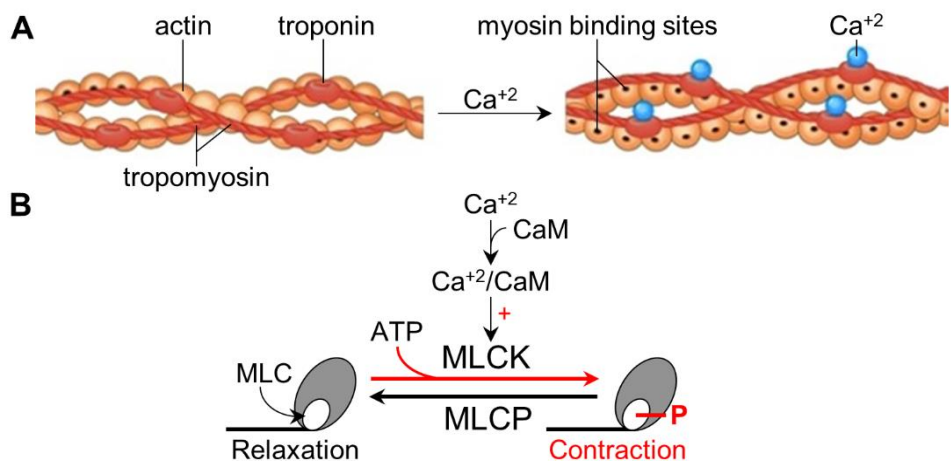

**Fig 6. Types of muscles and their contraction mechanisms.** (**A**) Striated muscles. When relaxed, the myosin binding sites, found on the actin filaments, are blocked by tropomyosin (**left**). $[Ca^{+2}]_{in}\uparrow$ exposes the myosin binding sites due to the binding of $Ca^{+2}$ to troponin, which induces conformational change of tropomyosin and leads to muscle contraction (**right**). (**B**) Non-striated muscles. When relaxed, a muscle-contraction regulatory enzyme, myosin-light-chain-kinase (MLCK), is inactive, and thus the myosin light chains (MLC) found at the base of the myosin heads (grey) are not phosphorylated (**left**). $[Ca^{+2}]_{in}\uparrow$ induces MLCK's activation, by binding to calmodulin (CaM), allowing the $Ca^{+2}$/CaM complex to bind to and activate MLCK. In order for enough MLC to be phosphorylated (red P) for allowing muscle contraction (**right**), a second enzyme that de-phosphorylates the MLC, myosin-light-chain-phosphatase (MLCP), should become inactive.

Pharyngeal muscles are classified as non-striated [28, 29]. Thus, a delayed inhibition of MLCP could explain the large delay between the early $[Ca^{+2}]_{in}\uparrow$ and the late contraction of the AI muscles.

**Proposed mechanism for the *delayed* contraction of the anterior isthmus: delayed inhibition of MLCP**

1. In the relaxed isthmus muscles, the activity-level of MLCP is high. Hence:

2. The early $[Ca^{+2}]_{in}\uparrow$ at the beginning of a pump is required, but is not sufficient for isthmus contraction. Inhibition of MLCP is also required.

3. We suggest that MCLP's inhibition occurs late during the pump, with a delay of ~75 ms at the anterior edge of the isthmus (Fig 7A).

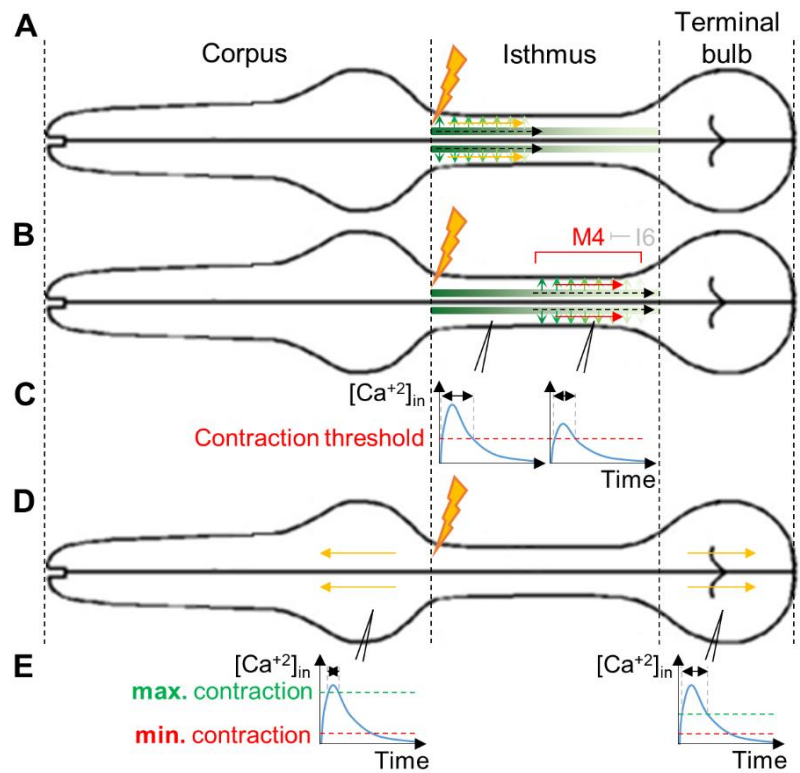

**Fig 7. Suggested mechanisms for the contraction dynamics of the various pharyngeal areas.** At the beginning of a pump, MC's firings at the corpus-isthmus border (lightning) locally depolarize two types of pharyngeal cells: the mc2 (green lines) and the pharyngeal muscles (white). DP spreads rapidly along the pharyngeal muscles, inducing a nearly-simultaneous $[Ca^{+2}]_{in}\uparrow$, and thus MLCK-activation early in the pump along the corpus, AI and TB muscles (solid yellow arrows). (**A-C**) Suggested mechanism for AI (**A**) and PI (**B**) dynamics: delayed and gradual inhibition of MLCP, combined with different MLCK-activation and $Ca^{+2}$ dynamics (**C**). (**A-B**) In the relaxed isthmus muscles, the activity-level of MLCP is high, and thus in addition to MLCK-activation, which is induced by DP at AI (**A**, solid yellow arrows) and by M4-activity at PI (**B**, solid red arrows), MLCP-inhibition is also required for muscle contraction. The inhibition of MLCP occurs gradually along these muscles due to a slow conductance of the electrical signal along the mc2 (dashed black arrows). DP of each point along the mc2 initiates a slow signal transduction pathway that inhibits MLCP at the adjacent muscle's segment in a delayed and local manner (green arrows), resulting in a delayed and gradual muscle contraction. (**B**) The pharyngeal neuron M4 is selectively inhibited, possibly by the pharyngeal neuron I6 (grey), during pumps in which the TB is not empty of food, resulting in lower-frequency of peristalsis relative to pumping. Only during pumps in which the TB is empty, is M4 free to fire, inducing a nearly-simultaneous $[Ca^{+2}]_{in}\uparrow$, and thus MLCK-activation along the entire PI prior to MLCP-inhibition. (**C**) Schematic $Ca^{+2}$ dynamics at the AI (left) and PI (right) muscles during a pump. $[Ca^{+2}]_{in}$ reach lower maximal values at the PI than at the AI. This results in shorter (longer) contractions of each PI (AI) muscle's segment, i.e., peristalsis (pumping). (**D-E**) Suggested mechanism for corpus and TB dynamics: rapid and nearly-simultaneous activation of MLCK along the entire muscles, and different $[Ca^{+2}]_{in}$-thresholds of contraction. (**D**) In the relaxed corpus and TB muscles, the activity-level of MLCP is low, and thus MLCK-activation alone is required and sufficient for muscle contraction. The rapidly-spreading DP, which induces early and nearly-

simultaneous [Ca$^{+2}$]$_{in}$↑, and thus MLCK-activation along the muscles (solid yellow arrows), results in early contraction-onset of the entire muscles. (**E**) Schematic Ca$^{+2}$ dynamics at the corpus (left) and TB (right) muscles during a pump. The slow (fast) progression of contraction-strength of the corpus (TB) muscles results from a high (low) [Ca$^{+2}$]$_{in}$-threshold of maximal muscle contraction: both groups of muscles have a similar minimal-contraction threshold, i.e., contraction-onset (dashed red lines), but have different maximal-contraction thresholds (dashed green lines). A high maximal-contraction threshold in the corpus results in contraction-strength that is proportional to [Ca$^{+2}$]$_{in}$, reaching its maximum at the end of the pump, while a low threshold in the TB results in reaching maximal contraction early in the pump. The high (low) maximal-contraction threshold also results in an early (late) muscle relaxation following its repolarization (which occurs near the peak of the graphs).

Such a slow inhibition accords with the time constants reported in the literature for non-striated muscles, where MLCP-inhibition, as well as MLCK-activation, are slow processes that occur via signal transduction pathways [30, 31]. The time from activating the transmembrane receptors to affecting the end-target enzymes can take from tens of milliseconds up to many minutes, fitting a slow inhibition process of ~75 ms of MLCP.

This proposed mechanism for the delayed isthmus contraction explains the slow spreading of the contraction as well: [Ca$^{+2}$]$_{in}$↑ occurs early at the beginning of a pump, and seems to occur rapidly throughout the entire AI (see hypothesis #2 for *gradual* contraction), thus activating MLCK nearly simultaneously along this entire region at the beginning of a pump. Hence, according to the above mechanism, only a gradual inhibition of MCLP along the AI could explain its gradual contraction, as elaborated in the following sections.

**Hypothesis #3 for the *gradual* contraction of the anterior isthmus: gradual inhibition of MLCP along the anterior isthmus muscles**

In order for MLCP to be inhibited gradually along the muscle, three conditions should be fulfilled regarding an external effector that signals the muscle's MLCP: it should (a) stretch along the entire AI, (b) relay the signal slowly, and (c) affect the activity of MLCP in a local manner, influencing small muscle's segments separately (Fig 5C).

Again, as in hypothesis #1 for *gradual* contraction, two types of cells may fulfil these requirements:

(1) Pharyngeal neurons: as mentioned, anatomically, several pharyngeal neurons stretch along the entire isthmus. Among these, two are good candidates for isthmus innervation, the M2's and the NSM's pairs, since only they form many synapses throughout the isthmus. Theoretically, each synapse could affect its adjacent muscle's segment separately, in a local manner [11, 32]. Functionally, the M2's appear to be better candidates, for two reasons: (a) manipulating M2's activity affects pumping, where their activation (ablation) increases (decreases) pumping rate, while manipulating NSM's activity has little effect on pumping [3, 12, 33]; (b) in a *C. elegans* ancestor, the M2's exclusively control AI motion, inducing peristalsis [4]. Although in *C. elegans* the AI exhibits pumping rather than peristalsis, preserved anatomical connections could result in a similar effect of controlling the motion of the same pharyngeal area.

However, killing either the M2's or NSM's still results in pharyngeal pumping of all pumping areas, including the AI [although pumping rate decreases; 3, 17, 33]. Thus, in *C. elegans*, the M2's are not exclusively required for AI motion. In addition, in order for the M2's to gradually inhibit MLCP along the AI, their conductance velocity should be (11.93 to 17.9 µm)/(~85 ms) = 0.14-0.21 µm/ms, which is much slower than the reported values for *C. elegans* neurons [7.1-35.3 µm/ms; 18].

(2) Marginal cells: as mentioned, anatomically, the mc2 stretch along the entire isthmus. In addition, gap junctions that spread throughout their entire length, connecting them to the isthmus muscles, could theoretically affect small muscle segments locally, as required [1]. Functionally, the mc2 are strongly innervated at their anterior edge by the exact same neurons that trigger each pump – the MC's neurons [11-13]. Thus, if the mc2 could *conduct* the electrical signal generated at their anterior end, and conduct it *slowly*, then they could have accounted for the gradual inhibition of the AI muscles.

Not much experimental data testing the possible function of the marginal cells in *C. elegans* pharynx exist. However, it was previously speculated that the marginal cells, or at least the mc2, could conduct electrical activity: their strong innervation by the MC's and the abundant gap junctions between them and the isthmus muscles make no sense if these cells cannot conduct [16]. In addition, while the large diameter of the marginal cells implies fast conductance, the abundant gap junctions along them may have an even greater effect, highly increasing their membrane permeability and slowing down their conductance speed [see PhaFIG 7A in 1].

**Proposed mechanism for the *gradual* contraction of the anterior isthmus: gradual inhibition of MLCP**

1. At the beginning of a pump, the MC's neurons: (a) innervate the mc2 at their anterior edge [11], and (b) rapidly depolarize the pharyngeal muscles, starting at the corpus-isthmus border [14, 21, 34].

    The early and nearly-simultaneous muscles' DP induces early and nearly-simultaneous $[Ca^{+2}]_{in}\uparrow$, and thus MLCK's activation along the entire isthmus muscle. We suggest that in the isthmus muscles, activation of MLCK is required but is not sufficient for muscle contraction, since MLCP should also be inhibited.

2. We suggest that the electrical signal generated at the anterior end of the mc2 propagates slowly along them. The arrival of the signal at each point along the mc2 initiates a slow signal transduction pathway that locally inhibits MLCP after tens of milliseconds at the adjacent muscle's segment, thus triggering a delayed and local muscle contraction. Note that either MLCP-inhibitor(s) or any upstream messenger(s) could diffuse through the mc2-pm5 gap junctions, since gap junctions allow the diffusion of small molecules, including second messengers [35].

3. The slowly propagating signal along the mc2 induces a gradual muscle contraction (Fig 7A).

Importantly, the suggested mechanism for the delayed and gradual isthmus contraction include several assumptions, which can be tested by appropriate experiments in the nematode. Conduction ability and velocity of the mc2 can be tested by intracellular measurements from these cells, and the involvement of MLCP-inhibition can be tested by measurements of MLCP/MLCP-inhibitors activity-level in relaxed vs. pumping pharynxes.

## Proposed mechanisms for the contraction dynamics of the other pharyngeal areas

### Posterior isthmus

The AI and PI have similar contraction dynamics: both start to contract late during a pump, and both contract slowly, in an anterior-to-posterior wave [3, 7].

Since the mc2 stretch throughout the entire isthmus, the same mechanisms suggested for the AI can also explain the contraction dynamics of the PI. Moreover, the most posterior (anterior) part of the AI (PI) starts contracting ~75+85=160 (~150) ms after pump-onset, implying a single continuous contraction process along the entire isthmus [see Fig 3 in 2, 7].

While the suggested mechanism for AI contraction explains the delayed and gradual contraction dynamics of the PI, it should additionally accord with several more observations characterizing the PI only. We thus start by describing these, and then describe an expanded mechanism that addresses all known facts.

Contraction dynamics unique to the PI:

(1) PI peristalsis is selectively coupled to the preceding pharyngeal pump [2, 10]:

   (i) Peristalsis does not occur in the absence of pumping.

   (ii) During pumping, peristalsis:pumping ratio is 1:3.4 on average.

   (iii) In pumps that are followed by peristalsis, peristalsis always starts after a constant interval of ~150 ms.

(2) M4 is necessary and sufficient for PI peristalsis:

   (i) Anatomically, M4 synapses only onto pharyngeal muscles in the isthmus and TB, where in the isthmus it synapses onto the entire PI [11, 26].

   (ii) Functionally:

      o When M4 is killed, peristalsis stops [17, 36].

      o M4's activation causes 97% of pumps to be followed by peristalsis [12].

(3) Each peristalsis and $[Ca^{+2}]_{in}\uparrow$ at the PI correlate well [27]:

   (i) Not every pump is followed by $[Ca^{+2}]_{in}\uparrow$ at the PI.

   (ii) Whenever peristalsis occurs, a significant increase in $Ca^{+2}$ at the PI is detected.

The above observations lead to the following conclusions: (a) both pumping and M4 are essential for PI peristalsis (based on points (1-i) and (2-ii)); (b) M4 induces $[Ca^{+2}]_{in}\uparrow$ (based on points (2-ii) and (3)).

Song and Avery [2] already reached these conclusions, and suggested that in peristalsis-induction, pumping represents PI DP → $[Ca^{+2}]_{in}\uparrow$. Namely, the researchers suggested that both pumping and M4 are essential for $[Ca^{+2}]_{in}\uparrow$ at the PI for inducing peristalsis. In contrast, we suggest a different explanation for the role of pumping in inducing peristalsis. We hypothesize that pumping represents MC's firings → mc2 activation → MLCP inhibition.

Thus, we propose for PI peristalsis a very similar mechanism to the one suggested earlier in this paper for the AI, with the only difference being the trigger for $[Ca^{+2}]_{in}\uparrow$: while at the AI it is DP, at the PI it is possibly M4's activity. As already mentioned, this mechanism explains the delayed and gradual contraction of the PI, but it does not address the following questions:

1. Why doesn't DP induce $[Ca^{+2}]_{in}\uparrow$ at the PI, in contrast to all other pharyngeal areas?
2. Why is the frequency of peristalsis lower than that of pumping?
3. Why does only the PI exhibit peristalsis, whereas all other pharyngeal areas exhibit pumping?

We suggest the following explanations:

1. We have no good explanation for this, and it thus remains an open question.
2. Since $[Ca^{+2}]_{in}\uparrow$ occurs exactly after the pumps that are followed by PI peristalsis, and from the conclusion that M4 induces $[Ca^{+2}]_{in}\uparrow$, the lower peristalsis-frequency question can be rephrased as follows: what determines the selective activity of M4?

   Since M4 does not have mechanosensory receptors facing the pharyngeal lumen it cannot directly sense the right conditions or timing for inducing contraction [11]. Sensory input should arrive from other neuron(s) that innervate M4, and which receive, directly or indirectly, sensory input from the pharyngeal lumen. The selective activity of M4 could result from either its selective inhibition during pumps not followed by peristalsis, combined with its consistent excitation or selective excitation during the complementary pumps, or from selective activation of M4 during pumps that are followed by peristalsis.

Out of the only two neurons that synapse on M4, I5 and I6 [11], we suggest that the selective activity of M4 results from its selective inhibition by I6. I6 applies chemical synapses onto the anterior end of M4, and has mechanoreceptors facing the lumen of the TB [11]. Thus, I6 could selectively inhibit M4 according to the TB's state: during pumps in which the TB is filled with food, I6's mechanoreceptors would stretch, which would activate I6. The activated I6 would inhibit M4, which would not be able to induce $[Ca^{+2}]_{in}\uparrow$ and thus peristalsis would not occur. In contrast, during pumps in which the TB is empty, I6 would not fire, and the non-inhibited M4 would be free to induce $[Ca^{+2}]_{in}\uparrow$, resulting in PI peristalsis. Thus, during pumps in which the TB is empty (filled with food), food would (not) be transferred from the AI to the TB, as desired.

This mechanism was already suggested by Avery [5], with a slight difference – that I5 would inhibit M4 rather than I6. However, while I5 has mechanoreceptors facing the lumen of the TB [11], strongly synapses onto M4 [11], and can probably inhibit M4 (I5 is possibly a glutamatergic neuron [37], and M4 expresses GLR-8 glutamate receptors [38]), I5 seems to fire during every pump [5] and thus cannot selectively inhibit M4.

We do not have a suggested candidate neuron for M4's activation (neither consistent nor selective) that accords with all available data: we disqualify I5 based on the results of Song and Avery [2], who show that M4 triggers PI peristalsis via the activation of a specific serotonin-receptor expressed on M4 – SER-7. However, I5 does not express serotonin-biosynthesis genes, and is thus probably not the M4-activating neuron [39]. In contrast, out of the pharyngeal neurons, only the NSM's pair express the serotonin-synthesizing genes, and have mechanoreceptors at the corpus-isthmus border [11, 40, 41]. While the NSM's do not synapse directly onto M4, they could affect M4 neurohumorally, by secreting serotonin to the pseudocoelomic fluid [3, 11]. However, NSM-ablated worms still exhibit PI peristalsis, suggesting that the NSM's do not play any important role in this behaviour [3, 33].

3. This question can also be rephrased as follows: AI pumping and PI peristalsis actually differ mainly in their time of relaxation: whereas at the AI increasingly more segments gradually join the contraction and relax together at the end of a pump, each of the PI' segments relaxes shortly after it starts contracting. Thus, this question can be rephrased as: what induces the fast relaxation of the PI muscles?

We suggest that while the (entire) isthmus muscles have a high $[Ca^{+2}]_{in}$ threshold for contraction-onset, the absolute elevation of $[Ca^{+2}]_{in}$ is smaller at the PI, resulting in shorter contractions of the PI segments (Fig 7C). This idea is supported by Shimozono, Fukano [27], but only during low pumping rates [see Figs 2B, 3B in 27]. Due to the poor time resolution of their technique, it is possible that the measurements during high pumping rates are less accurate.

**Proposed mechanism for the contraction dynamics of the posterior isthmus**

1. At the beginning of each pump, MC's firings generate an electrical signal at the anterior edge of the mc2, which propagates slowly along the mc2, inhibiting MLCP late during the pump and in a gradual manner along the PI.
2. M4's activation induces $[Ca^{+2}]_{in}\uparrow$ → MLCK activation. We suggest that in the isthmus muscles, MLCK-activation is required, but is not sufficient for muscle contraction, where MLCP should be inhibited as well. Moreover, we suggest that MLCK-activation occurs earlier during the pump than MLCP-inhibition, so that MLCP-inhibition time is the determining factor of contraction-onset time (Fig 7B).

3. The selective activity of M4, perhaps via selective inhibition by I6, stands for the lower frequency of peristalsis.

4. We suggest that a lower $[Ca^{+2}]_{in}\uparrow$ at the PI than at the AI results in peristalsis rather than pumping (Fig 7C).

**Corpus and terminal bulb**

The suggested mechanism for explaining the slow progression of contraction-strength of the corpus relative to the TB muscles should accord with the following findings [3, 6-8, 42]:

(1) The corpus muscles, as well as the TB muscles, contract in synchrony along their entire length.

(2) Corpus contraction is gradual, starting from a weak contraction at the beginning of a pump and gradually strengthening until reaching the maximal contraction at the end of a pump.

(3) TB contraction is rapid, reaching the maximal contraction very early during a pump, and remaining at this maximum throughout the pump.

(4) The corpus muscles repolarize and relax almost simultaneously, although relaxation is not quite complete. Then the TB muscles repolarize, about 50-60 ms after corpus relaxation-onset time, but are still contracted until they relax.

**Proposed mechanism for the contraction dynamics of the corpus and terminal bulb**

1. We hypothesize that in the relaxed corpus and TB muscles, in contrast to the isthmus muscles, MCLP is not active, resulting in early and nearly-simultaneous contraction of the entire muscle, that depends solely on the rapid and nearly-simultaneous DP → $[Ca^{+2}]_{in}\uparrow$ → MLCK-activation (Fig 7D).

2. The slow (fast) contraction of the corpus (TB) muscles could result from a high (low) $[Ca^{+2}]_{in}$-threshold of muscle's maximal contraction (Fig 7E): assuming a similar minimal-contraction $[Ca^{+2}]_{in}$-threshold in both groups of muscles, which would result in similar contraction-onset times, a high maximal-contraction $[Ca^{+2}]_{in}$-threshold in the corpus would result in contraction-strength that is proportional to $[Ca^{+2}]_{in}$, reaching its maximum at the end of DP, i.e., at the end of a pump, while a low threshold in the TB would result in reaching the maximal contraction-strength early in the pump.

3. The high (low) maximal-contraction $[Ca^{+2}]_{in}$-threshold of the corpus (TB) muscles must result in faster (slower) muscle relaxation following its repolarization, which accords with point (4) above (Fig 7E).

## Discussion

The pharynx of the nematode *C. elegans* has been studied for over four decades. Out of the two typical pharyngeal motions, pumping and peristalsis, the former is far more extensively studied. However, in most studies, little consideration is given to the fact that different pharyngeal areas exhibit slightly different dynamics; pumping is usually studied in the context of pumping-induction or regulation, where the motion of all pumping areas is represented by the motion of the grinder of the terminal bulb. Accordingly, not many proposed mechanisms exist for explaining how the various pharyngeal areas generate different dynamics. While several mechanisms exist, when inspected in detail they seem to be partial, either not adequately matching all relevant data, or being quite general, not explaining the observed phenomena in sufficient depth [2, 5, 10].

In this study, we propose possible mechanisms for explaining how diverse contraction and relaxation dynamics observed in the various pharyngeal areas: corpus, anterior isthmus (AI), posterior isthmus (PI), and terminal bulb (TB), are generated. In contrast to other proposed mechanisms, which address some of the pharyngeal areas, our suggested mechanisms deal with them all, explaining all observed dynamics in a way that contradicts none of the relevant data, whilst is also specific and detailed, explaining the phenomena to their very bottom.

In addition, our analyses propose a new function for pharyngeal cells, whose role in the *C. elegans* pharynx is thought to be primarily structural – to supply reinforcing strength to this muscular organ [1]. We hypothesize that the marginal cells 2 (mc2) are necessary for the generation of the isthmus motions – AI pumping and PI peristalsis. Still, it is important to note that our mechanisms do not propose a specific role for the other marginal cells in the pharynx – mc1 and mc3. Furthermore, we hypothesize that these cannot fulfil a similar role to that of the mc2: if the contraction of the corpus (TB) depended on the activity of the mc1 (mc3), i.e., on MLCP-inhibition, then according to our hypothesis these muscles could not have contracted so early in the pump. Their contraction-onset delay would have been several tens of milliseconds rather than the few milliseconds measured in the nematode [3].

While our analyses address many open questions, several aspects still need to be addressed. First, we did not analyse the motion of several pharyngeal parts: (1) the pm1-2, whose motions are far less studied and characterized; (2) the anterior tip of the corpus, which relaxes before the remainder of the corpus [8, 43], and whose motion is also far less studied; its early relaxation may involve the selective relaxation of a sub-compartment of the pm3, possibly induced by the M1 neuron [8, 11, 43]; (3) the longitudinally-, rather than radially-, oriented filaments of the pm6-7, which generate longitudinal motions of the grinder [1]; we consider only motions that result from the radially-oriented filaments, which open the pharyngeal lumen.

Second, we did not analyse the data collected for most pharyngeal neurons and did not integrate them into our proposed mechanisms. As suggested by Avery [5], the primary function of the pharyngeal nervous system is probably only of sensory nature: rather than directly producing pharyngeal motions, it allows these to change in response to a changing environment. In accordance with this suggestion, many neurons were found to play some role in regulating the pharyngeal motions, but are not necessary for their generation; e.g., I1's, I2's, M1, M2's, and NSM's regulate pumping-rate [12, 14, 33, 43, 44], M1 induce "spitting" [43], and I5 and M3's regulate pump-duration [5, 9, 45]. Only neurons whose activity is necessary for the generation of pharyngeal motions were analysed and integrated into our suggested mechanisms: the MC's pair, which are the major neurons that induce pharyngeal pumping, and M4, which is the only neuron required for PI peristalsis.

Finally, two questions remain unanswered in our analyses: (1) Why does DP not induce $[Ca^{+2}]_{in}\uparrow$ at the PI, in contrast to all other pharyngeal areas? (2) Which pharyngeal neurons activate M4 during pumps that are followed by peristalsis?

Appropriate experiments on *C. elegans* can be set up for testing the proposed electrical features and role of the mc2, as well as several more assumptions made here for the different proposed mechanisms; e.g., nearly-simultaneous $[Ca^{+2}]_{in}\uparrow$ along the entire AI, high (low) activity-level of MLCP in the relaxed isthmus (corpus and TB) muscles, lower $[Ca^{+2}]_{in}\uparrow$ at PI than at AI, higher maximal-contraction $[Ca^{+2}]_{in}$-threshold of corpus than TB muscles. Such experiments could also help shed light on some of the open questions raised by our analyses.

In a broader scope, this study, as well as many other studies of the *C. elegans* pharynx, may contribute to our understanding of the mammalian digestive system. Several similarities exist between these two systems: (1) both are composed of a syncytium of smooth muscle cells; (2) both exhibit coordinated motions, including peristalsis, which result from the coordinated contractions of the many smooth muscle cells that compose them; (3) both exhibit motions regulated by neurotransmitters and neuropeptides; and (4) both have excitation-contraction coupling that is triggered by $[Ca^{+2}]_{in}\uparrow$ via transmembrane 'L-type' voltage-dependent $Ca^{2+}$ channels.

While the general characteristics of the mammalian gastrointestinal smooth muscle cells are more or less uniform, there is a considerable diversity among different species and different regions of the gastrointestinal tract: in the mechanisms that regulate smooth muscle cell excitability, in the excitation-contraction coupling, and in the generation and regulation of the contractile force [46]. None of these differences has been fully explained. Studying *C. elegans* pharynx may aid in addressing many of these questions. For example, it could aid in explaining the generation of different contractile forces in different regions along the digestive tract. This exact question, but regarding the *C. elegans* pharynx, is addressed in our paper. Our hypotheses include differences in the composition of transmembrane ion-channel in the various muscle cells (and specifically, in the composition of the L-type $Ca^{2+}$ channels, i.e., EGL-19 channels), different $Ca^{+2}$ dynamics, and different basal level of activity of regulatory enzymes that affect smooth muscle contraction.

Our hypotheses can be tested by appropriate experiments in the gastrointestinal tract of mammals (a considerable diversity in the expression of ionic conductances was found in human gastrointestinal smooth muscle cells; however, this subject has not been studied in enough depth to clearly describe the resulting differences in contraction). In addition, the various contractile forces could result from variations in the composition of the contractile elements – actin and myosin. Specifically, in both human and *C. elegans* digestive systems, several isoforms of myosin heavy chains are distributed differently in muscle cells located at different regions along the digestive tract [47, 48]. In both species, the functional importance of this cellular heterogeneity is not yet understood. In this paper, we analyse the fundamental differences between the motion dynamics of the isthmus and of the other pharyngeal areas – corpus and TB. One isoform of the myosin heavy chains in *C. elegans*, *myo-1*, is distributed unevenly along the pharynx, being expressed in the corpus and TB but not in the isthmus muscles [47], and could thus possibly contribute to explaining the difference in the contraction dynamics of these regions. Studying this possible effect in *C. elegans* might shed light on human gastrointestinal muscles as well.

Studying *C. elegans* pharynx may also aid in explaining responses to contraction regulation in the mammalian digestive tract: much is known about the regulation of gastrointestinal muscle contraction by neurotransmitters, but much less is known about its regulation by neuropeptides [46]. In contrast, several neuropeptides, released from pharyngeal and extra-pharyngeal neurons, regulate pharyngeal pumping, and multiple mechanisms for modulation of pumping rate are known [49-52]. Such mechanisms could provide possible explanations for the regulation of human gastrointestinal cells as well, and can then be tested by appropriate experiments.

## Funding

This work was supported by two grants from the Israel Science Foundation: a F.I.R.S.T. (BIKURA) Grant (No. 193/17), and a Personal Research Grant (No. 902/16).